\documentclass[a4paper]{article}

\usepackage{INTERSPEECH2022}
\usepackage{cite,bm}
\usepackage{amssymb}
\usepackage{amsfonts}
\usepackage{comment}
\usepackage{amsmath,mathtools,amssymb, amsfonts}
\usepackage{algorithm}
\usepackage[noend]{algpseudocode}
\usepackage{dsfont}
\usepackage{booktabs}
\usepackage{flushend}
\usepackage[binary-units]{siunitx}
\usepackage{xcolor}
\usepackage{physics}
\pdfoutput=1

\AtBeginDocument{
  \abovedisplayskip     =0.5\abovedisplayskip
  \abovedisplayshortskip=0.5\abovedisplayshortskip
  \belowdisplayskip     =0.5\belowdisplayskip
  \belowdisplayshortskip=0.5\belowdisplayshortskip
}

\def\C{{\mathbb C}}
\def\R{{\mathbb R}}
\def\g{\color[gray]{0.5}}
\newcommand{\argmin}{\mathop{\rm arg~min}\limits} 

\title{Spatial Loss for Unsupervised Multi-channel Source Separation}

\name{
    Kohei Saijo$^{1,2*}$, 
    Robin Scheibler$^2$
    \thanks{
        *This work was done during an internship at LINE Corporation.
    }
}
\address{
  $^1$Waseda University, Japan \;  $^2$LINE Corporation, Japan}
\email{saijo@pcl.cs.waseda.ac.jp}

\begin{document}

\maketitle
\begin{abstract}
We propose a \textit{spatial loss} for unsupervised multi-channel source separation.
The proposed loss exploits the duality of direction of arrival (DOA) and beamforming: the steering and beamforming vectors should be aligned for the target source, but orthogonal for interfering ones.
The spatial loss encourages consistency between the mixing and demixing systems from a classic DOA estimator and a neural separator, respectively.
With the proposed loss, we train the neural separators based on minimum variance distortionless response (MVDR) beamforming and independent vector analysis (IVA).
We also investigate the effectiveness of combining our spatial loss and a \textit{signal loss}, which uses the outputs of blind source separation as the reference.
We evaluate our proposed method on synthetic and recorded (LibriCSS) mixtures.
We find that the spatial loss is most effective to train IVA-based separators.
For the neural MVDR beamformer, it performs best when combined with a signal loss.
On synthetic mixtures, the proposed unsupervised loss leads to the same performance as a supervised loss in terms of word error rate.
On LibriCSS, we obtain close to state-of-the-art performance without any labeled training data.

\end{abstract}
\noindent\textbf{Index Terms}: unsupervised learning, spatial loss, direction-of-arrival, beamforming, blind source separation

\section{Introduction}
Speech recordings are routinely corrupted by interference and background noise. 
Source separation has been studied as a powerful tool to mitigate these problems in speech systems, e.g., automatic speech recognition (ASR).
On the one hand, blind source separation (BSS) such as independent component analysis (ICA)~\cite{ica}, independent vector analysis (IVA)~\cite{iva_kim,iva_hiroe}, and independent low-rank matrix analysis (ILRMA)~\cite{ilrma} have been an area of intense research.
On the other hand, supervised learning of deep neural networks (DNNs) for single-channel source separation has been eagerly investigated~\cite{pit,dc,tasnet}.
In the multi-channel setup, methods that exploit a single-channel separation network to estimate spatial cues such as DNN-based minimum variance distortionless response (MVDR) beamforming~\cite{mvdr_heymann,mvdr_ochiai,cisdr} or BSS with a neural source model~\cite{duong_dnn,idlma, auxiva-iss-dnn} have led to stunning improvements in performance.
Such linear filtering techniques have also been shown empirically to be better front-ends for ASR than non-linear ones such as single-channel separation~\cite{haeb2020far}.
However, supervised learning requires access to the massive amount of mixtures and corresponding ground-truth signals.
Because such a dataset of natural recordings cannot be obtained, many prior works rely on simulation instead~\cite{pra,haeb2020far}.

Recently, unsupervised source separation with a \textit{signal loss}, which uses the outputs of BSS as pseudo-targets instead of ground-truth clean signals, has been proposed~\cite{Drude-unsup, Togami-unsup-kld}.
In~\cite{Drude-unsup}, time-frequency (TF) masks estimated by a blind spatial clustering technique were used to train a deep clustering model.
\cite{Togami-unsup-kld} proposed a loss function that evaluates Kullback-Leibler Divergence (KLD) between the posterior probability density function of the separated signals of BSS and that of the DNN-based separator to avoid overfitting to the errors in the BSS outputs.
These unsupervised losses enforce consistency between the output of BSS and that of DNN-based separators.

In contrast, we propose a \textit{spatial loss} function for unsupervised multi-channel source separation that enforces consistency of the estimated spatial parameters.
We exploit the duality of direction of arrival (DOA) and beamforming: since linear separation relies on the assumption that sources are mixed linearly, the mixing and the beamforming matrix should be inverse of each other.
We train a neural separator so that the beamforming vector for a given source should have a large inner product with the corresponding steering vector, while being close to orthogonal to those of the other sources
The steering vectors are obtained by conventional DOA estimation with the multiple signal classification (MUSIC) algorithm~\cite{music}.
We also investigate the effectiveness of combining the spatial loss and the signal loss~\cite{Togami-unsup-kld, cisdr} .
We train two types of neural separators with our proposed loss function, auxiliary function-based IVA (AuxIVA)~\cite{auxiva_ip} with a neural source model~\cite{auxiva-iss-dnn}, and neural MVDR beamformer~\cite{cisdr}.
We evaluate our proposed method with the recorded dataset LibriCSS~\cite{libricss}.
Synthetic mixtures are also used for a more detailed analysis that requires the groundtruth.

The key contributions are summarized as follows.
1)~We propose a spatial loss function for unsupervised multi-channel source separation.
We show the superiority of the spatial loss over the signal loss.
2)~We conduct extensive experiments using both synthetic mixtures and real-world recordings, whereas most prior works only considered the former.
The proposed loss leads to the same performance as a supervised loss in terms of word error rate on synthetic mixtures.
On LibriCSS, the proposed method outperformed strong baselines~\cite{libricss_conformer} and obtains close to state-of-the-art performance~\cite{libricss_sota} with no data besides that available in LibriCSS, nor any ground-truth.

\section{Background}
\label{sec:background}

Assuming $N$ sources are captured by $M$ microphones, the observed signal in the short-time Fourier transform (STFT) domain is represented with a mixing matrix $\bm{A}_{f}\in\C^{{M}\times{N}}$ as,
\begin{align}
  \label{eqn:signal_model}
    \bm{x}_{f,t} = {\bm{A}_f}{\bm{s}_{f,t}} + {\bm{b}_{f,t}},
\end{align}
where $\bm{s}_{f,t}$ is the clean sources vector, $\bm{b}_{f,t}$ is the background noise, and ${f} = {1,\ldots,F}$ and ${t} = {1,\ldots,T}$ are the frequency bin and the time frame index. 
Linear source separation is the problem of estimating a demixing matrix $\bm{W}_{f} \in \C^{{N}\times{M}}$, that estimates the sources as,
\begin{align}
  \label{eqn:separation}
    \bm{s}_{f,t} \approx {\bm{W}_f}{\bm{x}_{f,t}}.
\end{align}
Thus, the optimal demixing matrix should satisfy ${\bm{W}_f}{\bm{A}_f} \approx\bm{I}$, where $\bm{I}$ is the identity matrix.
In the following, $^\top{}$ and $^{\mathsf{H}}$ denote the transpose and Hermitian transpose of vectors or matrices.

\begin{figure}[t]
\centering
\centerline{\includegraphics[width=0.55\linewidth]{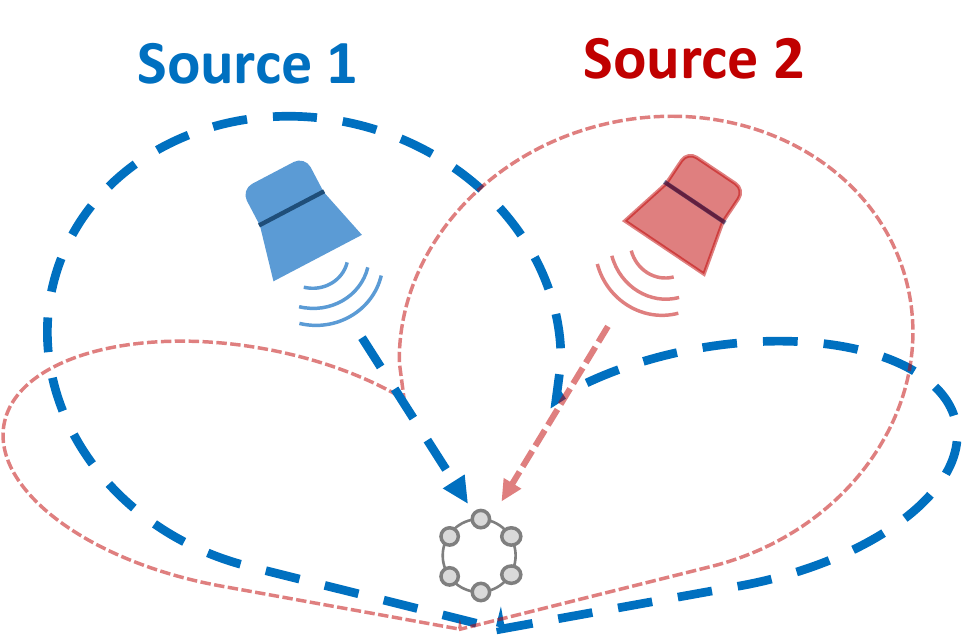}}
\vspace{-2mm}
\caption{Duality of beamforming and DOA. Beam is directed to target and null is directed to interference.}
\label{fig:beamforming}
\vspace{-3mm}
\end{figure}

\subsection{DOA Estimation}
\label{subsec:doa_estimation}
Knowing the location of each microphone, $\bm{d}_{m} \in \R^{3}$, the steering vector of the $n$th source is,
\begin{align}
  \label{eqn:stv}
    \bm{a}_{f}(\bm{q}_{n}) = M^{-\frac{1}{2}} \left[e^{j\omega_{f}\bm{d}_{1}^\top{} \bm{q}_{n}}, \ldots, e^{j\omega_{f}\bm{d}_{M}^\top{} \bm{q}_{n}} \right]^\top{},
\end{align}
where $\omega_{f}=2\pi\frac{f}{c}$ with sound speed $c$.
$\bm{q}_{n}\in\R^{3}$ is a unit-length vector pointing towards the $n$th source, and is represented with elevation $\phi$ and azimuth $\theta$ as $\bm{q}=[\rm{cos}\theta \rm{sin}\phi, \rm{sin}\theta \rm{sin}\phi, \rm{cos}\phi]^\top{}$.
Here we consider the MUSIC method~\cite{music}.
MUSIC assumes that $\bm{a}_{f}(\bm{q}_n)$ are orthogonal to the subspaces spanned by the $M-N$ least eigenvectors of the covariance matrix of (\ref{eqn:signal_model}), i.e., the noise subspace.
Let $\bm{E}_{f}$ be a matrix containing a basis for the noise subspace in its columns.
Then, $\bm{q}_{n}$ is estimated by finding local maxima of the following cost function,
\begin{align}
  \label{eqn:music_cost}
    \mathcal{G}(\bm{q}_{n}) = \frac{1}{F}\sum\nolimits_{f} (|\bm{E}_{f}^{\mathsf{H}} \bm{a}_{f}(\bm{q}_{n})|^2)^{-1}.
\end{align}
Hereafter, we denote $\bm{a}_{f}(\bm{q}_{n})$ as $\bm{a}_{n,f}$.


\subsection{Source Separation}
\label{subsec:separation}
Here we mention two separation methods, AuxIVA and MVDR.
\\
\textbf{AuxIVA} estimates the demixing matrix by likelihood maximization assuming the prior distribution of sources $p(y_{n,f,t})$, where $y_{n,f,t}$ is the $n$th separated signal.
The demixing matrix is estimated by minimizing the following negative log-likelihood,
\begin{align}
  \label{eqn:iva-cost}
    \mathcal{J}=-2T\sum_{f}\log|\mathrm{det}{\bm{W}_f}| + 
    \sum_{n,f,t} \frac{|\bm{w}^{\mathsf{H}}_{n,f}\bm{x}_{f,t}|^2}{r_{n,f,t}},
\end{align}
where $\bm{w}^{\mathsf{H}}_{n,f}$ is the $n$th row vector of $\bm{W}_{f}$ and $r_{n,f,t}$ the variance of the source prior $p(y_{n,f,t})$.
$\bm{W}_{f}$ can be updated with the techniques such as iterative projection~\cite{auxiva_ip} or iterative source steering (ISS)~\cite{iss}.
While conventional IVA uses a fixed source prior, replacing the source model with a trained DNN has recently demonstrated high performance~\cite{auxiva-iss-dnn}.
\\
\textbf{MVDR} computes the demixing vector that minimizes the noise variance under the distortionless constraint of the target source,
\begin{align}
  \label{eqn:mvdr_problem}
    {\bm{w}}_{n,f} = \argmin_{\bm{w}_{n,f}} {{{\bm{w}^{H}_{n,f}}}{\bm{R}^{(n)}_{n,f}}{\bm{w}}_{n,f}} ~~~s.t.~~~ {\bm{w}^{H}_{n,f}}{\tilde{\bm{a}}_{n,f}} = 1,
\end{align}
where ${\bm{R}^{(n)}_{n,f}}$ is the spatial covariance matrix of the noise.
The steering vector ${\tilde{\bm{a}}_{n,f}}$ and ${\bm{R}^{(n)}_{n,f}}$ can be obtained using DNNs~\cite{mvdr_heymann}.

\subsection{Unsupervised Learning using Signal-based Loss}
\label{subsec:sig_loss}
Recently, learning source separation using the separated signals from a BSS method as the pseudo-target signals has been proposed~\cite{Drude-unsup,Togami-unsup-kld}.
Let $\bm{\bar{y}}$ and $\bm{\hat{y}}$ be the separated signal of BSS and that of DNN-based separator.
The training is simply done with the loss between the two signals, which we call \textit{signal loss}, as 
\begin{align}
  \label{eqn:sig_loss}
    \mathcal{L}_{\rm{sig}} = \mathcal{L}(\bm{\bar{y}}, \bm{\hat{y}}),
\end{align}
where $\mathcal{L}$ is the permutation invariant loss.
Here we consider two loss functions, KLD~\cite{Togami-unsup-kld} and CI-SDR~\cite{cisdr}.

\noindent
\textbf{KLD}:
Assuming that each TF bin of the source follows a time-varying complex Gaussian distribution, 
i.e., $y_{n,f,t} \sim \mathcal{N}(0, r_{n,f,t})$,
KLD loss for a single TF bin is, 
\begin{gather}
  \label{eqn:kld_loss}
    \mathcal{L}_{\rm{kld}} =  \frac{|\hat{y}_{\pi(n),f,t}-\bar{y}_{n,f,t}|^2}{\hat{r}_{\pi(n),f,t}} + \frac{\bar{r}_{n,f,t}}{\hat{r}_{\pi(n),f,t}} + \mathrm{log}\frac{\hat{r}_{\pi(n),f,t}}{\bar{r}_{n,f,t}}  - 1,
\end{gather}
where $\pi(n)$ denotes the optimal assignment among all the permutations to minimize the loss. 
The final loss is obtained by summing (\ref{eqn:kld_loss}) over all TF bins and sources.

\noindent
\textbf{CI-SDR}:
Let $\bm{\bar{s}}$ and $\bm{\hat{s}}$ be the \textit{time-domain} separated signals of BSS and DNN.
Further define a matrix containing $K$ shifts of $\bm{\bar{s}}$ in its columns, $\bm{\tilde{S}}=[\bm{\tilde{s}}_{1},\dots,\bm{\tilde{s}}_{I}]^\top{}\in\mathbb{R}^{{I}\times{K}}$, with $i$th row $\bm{\tilde{s}}_{i}=[\bar{s}_{i-1},\dots,\bar{s}_{i-K}]^\top{} \in\mathbb{R}^{K}$ where $i=1,\dots,I$ is the time index.
Then, let $\bm{\alpha}_{n}=(\bm{\tilde{S}}_{n}^\top{}\bm{\tilde{S}}_{n})^{-1}\bm{\tilde{S}}_{n}^\top{}\bm{\hat{s}}_{\pi(n)}$, CI-SDR loss is,
\begin{align}
  \label{eqn:cisdr}
    \mathcal{L}_{\rm{sdr}} = \sum_{n} -10\log_{10}{\left(\frac{||\bm{\tilde{S}}_{n}\bm{\alpha}_{n}||^2}{||\bm{\tilde{S}}_{n}\bm{\alpha}_{n}-\hat{\bm{s}}_{\pi(n)}||^2}\right)}.
\end{align}

\section{Proposed Method}
\label{sec:prop}

\begin{figure}[t]
\centering
\centerline{\includegraphics[width=0.9\linewidth]{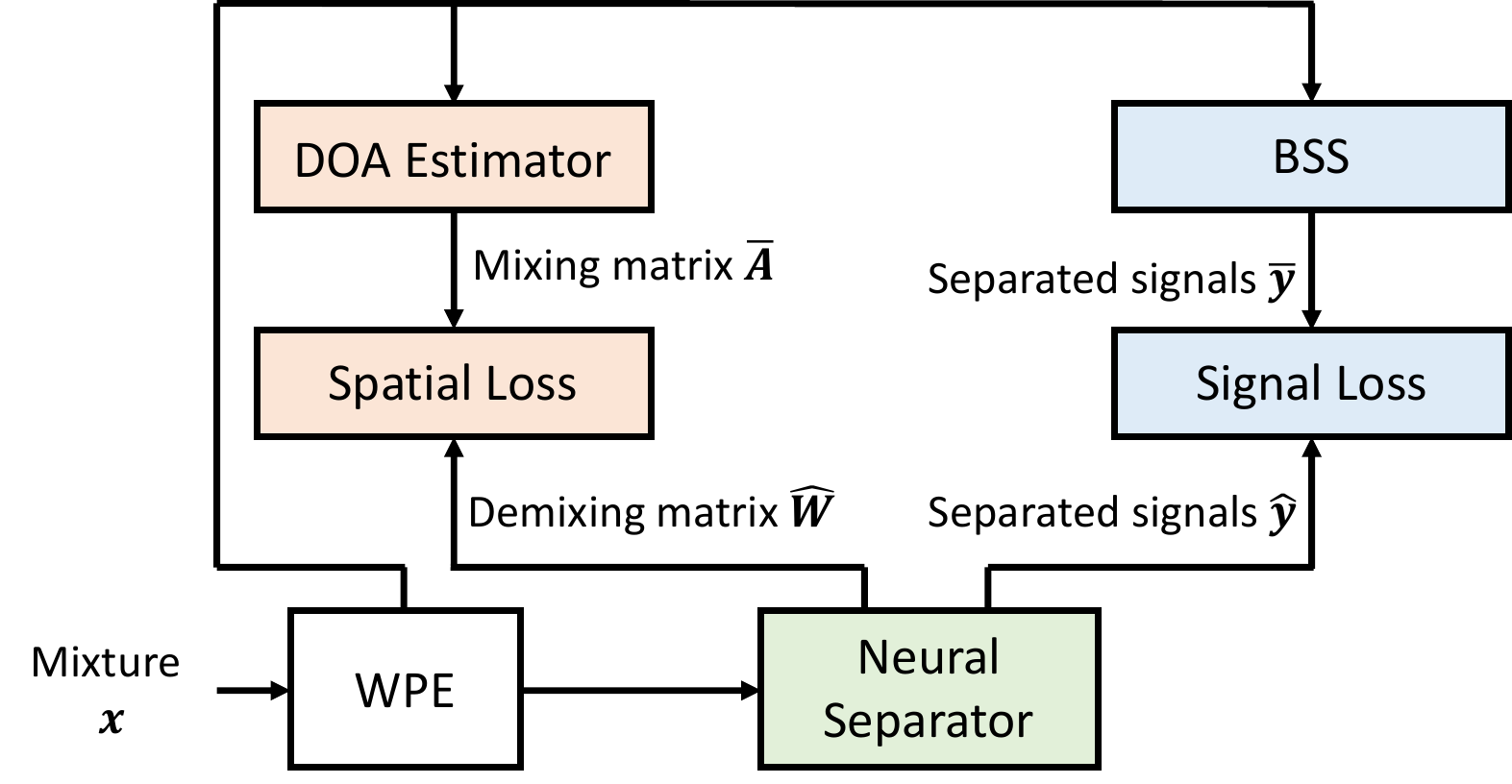}}
\vspace{-3mm}
\caption{Overview of our proposed method. We train a separator with blindly estimated speech signals and mixing matrices.
}
\label{fig:overview}
\vspace{-3mm}
\end{figure}

\begin{table*}[t]
\begin{center}
\caption{Average SDR in decibels, and STOI, PESQ, and WER of separated signals of IVA from the test set of WSJ1-mix dataset. Training is done with 3 channels, and evaluation is done with 2, 3 and 6 channels.  The proposed method is indicated by a star ($\star$).}
\label{table:iva_result_wsj}
\vspace{-3mm}
\scriptsize
\begin{tabular*}{\linewidth}{@{}ll@{\extracolsep{\fill}}rrr@{}}
\toprule
    {Algo.} 
    & {Loss} 
    & {\begin{tabular}{rrrr} \multicolumn{4}{c}{2 channels}\\{SDR$^\uparrow$} & {STOI$^\uparrow$} & {PESQ$^\uparrow$} & {WER$^\downarrow$}\end{tabular}}
    & {\begin{tabular}{rrrr} \multicolumn{4}{c}{3 channels}\\{SDR$^\uparrow$} & {STOI$^\uparrow$} & {PESQ$^\uparrow$} & {WER$^\downarrow$}\end{tabular}}
    & {\begin{tabular}{rrrr} \multicolumn{4}{c}{6 channels}\\{SDR$^\uparrow$} & {STOI$^\uparrow$} & {PESQ$^\uparrow$} & {WER$^\downarrow$}\end{tabular}}\\

\midrule
Unprocess.       & -             &\begin{tabular}{rrrr}-0.4~~~ &0.738~~ &1.213~~ &114.9\%\end{tabular}  &\begin{tabular}{rrrr}-0.4~~~ &0.738~~ &1.213~~ &114.9\%\end{tabular}  &\begin{tabular}{rrrr}-0.4~~~ &0.738~~ &1.213~~ &114.9\%\end{tabular} \\
\midrule
Gauss-IVA        & -             &\begin{tabular}{rrrr}8.8~~~  &0.865~~ &1.630~~~~ &40.6\% \end{tabular} &\begin{tabular}{rrrr}12.5~~~ &0.907~~ &1.920~~~~ &26.3\% \end{tabular} &\begin{tabular}{rrrr}16.8~~~ &0.937~~ &2.340~~~~ &21.2\% \end{tabular} \\
DNN-IVA          & KLD           &\begin{tabular}{rrrr}9.8~~~  &0.877~~ &1.666~~~~ &36.0\% \end{tabular} &\begin{tabular}{rrrr}13.4~~~ &0.917~~ &1.967~~~~ &22.5\%\end{tabular}  &\begin{tabular}{rrrr}16.9~~~ &0.941~~ &2.292~~~~ &19.0\% \end{tabular} \\
DNN-IVA          & SDR           &\begin{tabular}{rrrr}10.5~~~ &0.886~~ &1.701~~~~ &33.6\% \end{tabular} &\begin{tabular}{rrrr}14.2~~~ &0.923~~ &2.016~~~~ &21.0\% \end{tabular} &\begin{tabular}{rrrr}18.0~~~ &0.950~~ &2.408~~~~ &18.4\% \end{tabular} \\
DNN-IVA$\star$   & DOA$_{1}$     &\begin{tabular}{rrrr}10.4~~~ &0.888~~ &1.720~~~~ &31.5\% \end{tabular} &\begin{tabular}{rrrr}14.2~~~ &0.927~~ &2.056~~~~ &19.0\% \end{tabular} &\begin{tabular}{rrrr}\bf{19.4}~~~ &\bf{0.959}~~ &\bf{2.596}~~\,\, &\bf{12.7}\% \end{tabular} \\
DNN-IVA$\star$   & DOA$_{2}$     &\begin{tabular}{rrrr}\bf{11.2}~~~ &\bf{0.899}~~ &1.737~~~~ &28.7\% \end{tabular} &\begin{tabular}{rrrr}14.4~~~ &\bf{0.931}~~ &2.050~~\,\, &\bf{17.0}\% \end{tabular} &\begin{tabular}{rrrr}18.0~~~ &0.957~~ &2.471~~~~ &13.6\% \end{tabular} \\
DNN-IVA$\star$   & KLD+DOA$_{2}$ &\begin{tabular}{rrrr}11.0~~~ &0.894~~ &\bf{1.742}~~\,\, &\bf{27.9}\% \end{tabular} &\begin{tabular}{rrrr}\bf{14.8}~~~ &0.930~~ &\bf{2.091}~~~~ &17.3\% \end{tabular} &\begin{tabular}{rrrr}18.7~~~ &0.954~~ &2.522~~~~ &14.2\% \end{tabular} \\
DNN-IVA$\star$   & SDR+DOA$_{2}$ &\begin{tabular}{rrrr}10.7~~~ &0.890~~ &1.723~~~~ &30.0\% \end{tabular} &\begin{tabular}{rrrr}14.5~~~ &0.927~~ &2.056~~~~ &17.5\% \end{tabular} &\begin{tabular}{rrrr}18.6~~~ &0.954~~ &2.494~~~~ &13.5\% \end{tabular} \\
\g{DNN-IVA}      & \g{Sup.SDR}   &\begin{tabular}{rrrr}\g{11.4}~~~ &\g{0.898}~~ &\g{1.765}~~~~ & \g{29.3\%} \end{tabular} &\begin{tabular}{rrrr}\g{15.6}~~~ &\g{0.934}~~ &\g{2.133}~~~~ &\g{17.2\%} \end{tabular} &\begin{tabular}{rrrr}\g{20.4}~~~ &\g{0.960}~~ &\g{2.621}~~~~ & \g{13.1\%} \end{tabular} \\
\bottomrule

\end{tabular*}
\end{center}
\vspace{-7mm}
\end{table*}

We propose a \textit{spatial loss} function for unsupervised multi-channel source separation.
We exploit the duality of DOA and beamforming.
The beamforming vector for a given source should have a large inner product with the corresponding steering vector, while being close to orthogonal to those of competing sources.
DOAs are estimated by a MUSIC-based technique~\cite{doamm}.
Different from the signal-based approach described in Sec.\ref{subsec:sig_loss}, our proposed method explicitly imposes the spatial constraints on the demixing matrix.

\subsection{Spatial Loss Function}
\label{subsec:spatial_loss}
We propose a spatial loss function that exploits DOAs estimated by MUSIC as the pseudo-targets.
As shown in Fig.\ref{fig:beamforming}, there is a duality between beamforming and DOA, i.e., the beam should be directed to the target source direction and the null to the interference direction.
Let $\bm{\bar{A}}_{f}$ be the mixing matrix computed with the estimated DOAs and $\bm{\hat{W}}_{f}$ be the demixing matrix estimated by the neural separator.
Our proposed loss function is,
\begin{align}
  \label{eqn:doa_loss_matrix}
    \mathcal{L}_{\rm{doa}} = \min_{\bm{\Pi}} \sum\nolimits_{f} \mathds{1}^\top (|\bm{\Pi} - |\bm{\hat{W}_{f}}\bm{\bar{A}_{f}}||)\mathds{1},
\end{align}
where $|\bm{\hat{W}_{f}}\bm{\bar{A}_{f}}|$ is normalized so that all the elements range from 0 to 1.
$\mathds{1}$ is a vector whose all the elements are one and $\mathds{1}^\top \bm{X} \mathds{1}$ computes the sum of all the elements of a matrix $\bm{X}$.
$\bm{\Pi}$ denotes a permutation matrix, where only one element in each row and each column is 1 and the others are 0.
For example, if $N=2$, $\Pi$ is either $\smqty[1 & 0 \\ 0 & 1]$ or $\smqty[0 & 1 \\ 1 & 0]$.
Here we consider two options of normalizing $|\bm{\hat{W}_{f}}\bm{\bar{A}_{f}}|$.
One is to normalize rows and columns of $|\bm{\hat{W}_{f}}|$ and $|\bm{\bar{A}_{f}}|$, respectively, and the other is to normalize the rows of $|\bm{\hat{W}_{f}}\bm{\bar{A}_{f}}|$.
We denote the former as DOA$_1$ loss and the latter as DOA$_2$ loss, respectively.

\subsection{DOA Estimator}
\label{subsec:prop_doa_estimation}

We estimate DOA using the MUSIC with MM refinement algorithm~\cite{doamm}.
For highly overlapped case, we simply obtain $\bm{q}_{n}$ from the entire signal $\bm{x}$ with length $T$.
However, when the overlap is small, we found out that such way of estimation often missed the DOA of the less active source.
To bypass this problem, we estimate DOAs at regular intervals with sliding window and cluster them.
The detailed procedure is as follows.
\begin{enumerate}
    \setlength{\itemsep}{2pt}
    \setlength{\parskip}{0pt}
    \setlength{\leftskip}{-10pt}
    \item Obtain $N$ DOAs for each window of length $L$ and shift $S$.
    \item Group them into $K \geq N$ clusters $C_1,\ldots,C_K$ with $E_1,\ldots,E_K$ elements by two-dimentional k-means on elevation $\phi$ and azimuth $\theta$. The centroid of $C_k$, $(\phi_{k}, \theta_{k})$, represents the DOA of the $k$th source.
    \item Remove $C_k$ if $E_k$ is less than a threshold $E_{thres}$.
    \item Remove the cluster with the lower number of elements that satisfies $|\theta_{k}-\theta_{k'}|<\theta_{thres}$ ($1 \leq k < k' \leq K$). 
    \item Select top-$N$ DOAs with the largest $E_k$.
\end{enumerate}
Because sudden estimation errors occur at some time frames and they have large impact on clustering, we set $K$ more than $N$ in step~2, and remove such outliers in step~3.
In step~4, because the clusters with close centroids are considered to belong to the same source, we remove it.
Finally, we obtain $N' \leq N$ DOAs from the centroids of remaining cluster.
Since we define the spatial loss assuming the existence of $N$ sources, mixtures with $N'<N$ are not used for training.

The spatial loss requires knowledge of the microphone locations to obtain DOAs.
It is thus applicable to datasets where most samples are recorded with a few known devices.
We believe this to be reasonable.
For most applications of interest, the geometry of the microphone array is known, e.g., smart speakers and conferencing systems.
If it is unknown, but sufficient recordings from the same device are available, it can be estimated using blind calibration techniques~\cite{thrun_calibration, krekovic_calibration}.

\subsection{Neural Separator}
\label{subsec:separator}
As in Fig.~\ref{fig:overview}, we first apply weighted prediction error (WPE)~\cite{wpe}, and then the dereverberated mixture is separated by the neural separator.
We consider two neural separators, AuxIVA with a neural source model (DNN-IVA)~\cite{auxiva-iss-dnn} and DNN-based MVDR beamforming (DNN-MVDR)~\cite{mvdr_heymann, mvdr_ochiai, cisdr}.
In DNN-IVA, the DNN estimates $1/r_{n,f,t}$ in (\ref{eqn:iva-cost}) in the form of a TF mask, and spatial model update is done with ISS~\cite{iss}.
The network is composed of three gate linear units\cite{glu} and a transpose convolution layer as in~\cite{auxiva-iss-dnn}.
From~\cite{auxiva-iss-dnn}, the size of the intermediate feature is changed to 256, and group normalization~\cite{groupnorm} with four groups is used instead of batch normalization.
In DNN-MVDR, the DNN estimates TF masks of the target source and the noise for each source, which are used to compute the demixing matrix.
The network consists of three bi-directional long short-term memory layers and two feed forward layers as in~\cite{cisdr}, where we changed the number of units to 512.

\section{Experiments}
\label{sec:experiments}

\begin{table}[t]
\begin{center}
\caption{Average SDR in decibels, and STOI, PESQ, and WER of separated signals of MVDR from the test set of WSJ1-mix dataset.  The proposed method is indicated by a star ($\star$).}
\label{table:mvdr_result_wsj}
\vspace{-3mm}
\scriptsize
\begin{tabular}{@{}l@{\extracolsep{\fill}}llrrrr@{}}
\toprule
{Ch.} & {Algo.} & {Loss} & {SDR$^\uparrow$} & {STOI$^\uparrow$} & {PESQ$^\uparrow$} & {WER$^\downarrow$} \\

\midrule
6     & DNN-MVDR         & SDR           &\bf{17.1} &0.944 &2.333 &17.1\%  \\
      & DNN-MVDR$\star$  & DOA$_{2}$     &12.8 &0.906 &2.033 &25.9\%  \\
      & DNN-MVDR$\star$  & SDR+DOA$_{2}$ &\bf{17.1} &\bf{0.945} &\bf{2.337} &\bf{16.6}\%  \\
      & \g{DNN-MVDR}     & \g{Sup.SDR}   &\g{19.0} &\g{0.952} &\g{2.512} & \g{15.4\%} \\
      
\bottomrule

\end{tabular}
\end{center}
\vspace{-6mm}
\end{table}

\begin{table*}[t]
\begin{center}
\caption{WER of LibriCSS dataset. Sessions 1 to 9 were used for training. The proposed method is indicated by a star (${\star}$). Methods with a dagger ($\dag$) are initialized with weights pre-trained on WSJ1-mix. The others are trained from scratch. }
\label{table:result_libricss}
\vspace{-3mm}
\scriptsize
\begin{tabular*}{\linewidth}{@{}l@{\extracolsep{\fill}}lcrr@{}}
\toprule
    {Algo.} & {Loss} & {Ch.} 
    & {\begin{tabular}{rrrrrrr} \multicolumn{7}{c}{WER for session 0-9 (all) [\%]}\\{0S\;\;} & {0L\,} &{10.0} & {20.0} & {30.0} &{40.0} &{Avg}\end{tabular}}
    & {\begin{tabular}{rrrrrrr} \multicolumn{7}{c}{WER for session 0 (test) [\%]}\\{0S\;\;} & {0L\,} &{10.0} & {20.0} & {30.0} &{40.0} &{Avg}\end{tabular}}\\
\midrule
Unprocess. & -       & -    &\begin{tabular}{rrrrrrr}11.8  &11.7  &18.8  &27.2  &35.6  &43.3  &26.4 \end{tabular}  &\begin{tabular}{rrrrrrr}11.3\;\;  &7.3   &19.3  &25.9  &34.5  &38.2  &24.4  \end{tabular}\\

\midrule
\multicolumn{2}{c}{Chen \textit{et al.}~\cite{libricss_conformer}} &7 &\begin{tabular}{rrrrrrr}7.2\;\;   &7.5\;\,   &9.6  &11.3  &13.7  &15.1 &11.2 \end{tabular} &\begin{tabular}{rrrrrrr}--~~~~~   &--~~~~~   &--~~~~~  &--~~~~~  &--~~~~~  &--~~~~~ &--~~ \end{tabular} \\
\multicolumn{2}{c}{Wang \textit{et al.}~\cite{libricss_sota}} &7 &\begin{tabular}{rrrrrrr}5.8~~   &5.8~~   &5.9~\,  &6.5~~  &7.7~~ &8.3\;\; &6.8 \end{tabular} &\begin{tabular}{rrrrrrr}--~~~~~   &--~~~~~   &--~~~~~  &--~~~~~  &--~~~~~  &--~~~~~ &--~~ \end{tabular} \\

\midrule
DNN-IVA      & Sup. (WSJ1-mix)   &3 &\begin{tabular}{rrrrrrr}7.1\;\; &7.3\;\, &9.0  &11.3  &13.5  &14.4 &10.9 \end{tabular} &\begin{tabular}{rrrrrrr}6.6\;\; &6.5\;\, &9.7  &11.7 &10.7 &12.0\;\, &9.8 \end{tabular}\\
\midrule
Gauss-IVA    & -          &3 &\begin{tabular}{rrrrrrr}7.5\;\; &7.4\;\, &9.2  &12.1  &15.4  &18.8 &12.3 \end{tabular} &\begin{tabular}{rrrrrrr}8.6\;\; &5.4 &10.5 &11.7 &13.0 &13.0 &10.8 \end{tabular}\\
DNN-IVA      & KLD        &3 &\begin{tabular}{rrrrrrr}7.7\;\; &7.8\;\, &9.9  &12.8  &15.5  &17.0 &12.3 \end{tabular} &\begin{tabular}{rrrrrrr}8.4\;\; &5.9\;\, &9.6  &11.9 &12.9 &10.7 &10.2 \end{tabular}\\
DNN-IVA      & SDR        &3 &\begin{tabular}{rrrrrrr}8.2\;\; &8.0\;\, &9.5  &12.1  &15.2  &17.5 &12.3 \end{tabular} &\begin{tabular}{rrrrrrr}8.6\;\; &5.7     &10.0 &12.1 &12.5 &12.1 &10.5 \end{tabular}\\
DNN-IVA$\star$ & DOA$_1$ &3 &\begin{tabular}{rrrrrrr}7.1\;\; &7.1\;\, &9.4  &12.2  &15.0  &17.3 &11.9 \end{tabular} &\begin{tabular}{rrrrrrr}6.4\;\; &\bf{5.2}\;\, &9.6  &10.8 &14.0 &11.5 &10.0 \end{tabular}\\
DNN-IVA$\star$ & DOA$_2$ &3 &\begin{tabular}{rrrrrrr}7.1\;\; &7.3\;\, &\bf{8.8}  &\bf{10.3}  &13.4  &15.1 &10.7 \end{tabular} &\begin{tabular}{rrrrrrr}6.6\;\; &5.5\;\, &9.4  &10.0 &11.9 &10.7\;\, &9.3 \end{tabular}\\
DNN-IVA$\star$ & KLD+DOA$_2$    &3 &\begin{tabular}{rrrrrrr}7.2\;\; &7.3\;\, &8.9  &10.4  &13.8  &14.8 & 10.8\end{tabular} &\begin{tabular}{rrrrrrr}6.3\;\; &5.4\;\; &9.2\;\,  &\bf{9.7}  &11.8 &10.1\;\, &9.0 \end{tabular}\\ 
DNN-IVA${\star\dag}$ & DOA$_2$ &3 &\begin{tabular}{rrrrrrr}\bf{6.5}\;\; &\bf{6.9}\;\, &9.1  &10.8  &\bf{13.0}  &\bf{14.1} &\bf{10.5} \end{tabular} &\begin{tabular}{rrrrrrr}\bf{5.9}\;\; &5.6\;\; &\bf{9.0}\;\,  &\bf{9.7}\;\,  &\bf{9.9}\;\;  &\bf{9.1}\;\, &\bf{8.4}  \end{tabular}\\
DNN-IVA${\star\dag}$ & KLD+DOA$_2$  &3 &\begin{tabular}{rrrrrrr}6.8\;\; &7.2\;\, &9.8  &12.7  &14.2  &14.8 &11.4 \end{tabular} &\begin{tabular}{rrrrrrr}6.1\;\, &5.3\; &10.9\; &9.9  &12.7  &11.3\;\, &9.7 \end{tabular}\\
\bottomrule

\end{tabular*}
\end{center}
\vspace{-7mm}
\end{table*}

\subsection{Datasets and Experimental Setup}
\label{subsec:setup}
\textbf{WSJ1-mix}:
We used six channel synthetic mixtures to evaluate the separation performance with speech metrics using the ground-truth.
It consisted of speech from the WSJ1 corpus~\cite{wsj1} and noise from the CHIME3 dataset~\cite{chime3} sampled in 16~kHz.
The reverberation times were chosen randomly from \SIrange{200}{600}{\milli\second}.
The number of sources was two, with relative power from \SIrange{-5}{5}{\decibel}.
The noise was scaled to attain an SNR between \SIrange{10}{30}{\decibel}.
Training, validation, and test sets contained 37416, 503, and 333 mixtures, approximately 98.5, 1.33 and 0.85 hours of mixtures, respectively.
To evaluate WER, we trained an ASR system with clean anechoic signals using the \textsf{wsj/asr1} recipe from the ESPNet framework~\cite{espnet}.
WER for the clean, anechoic test set was \SI{9.25}{\percent}.

During training, the batch size was 16 and the input signal length was 7 seconds. 
Network parameters were optimized using the Adam optimizer~\cite{adam} with learning rate $10^{-5}$.
When combining signal and spatial loss, we took weighted sum of them, i.e., $\mathcal{L}_{\rm{doa}}+\alpha \mathcal{L}_{\rm{sig}}$, where $\alpha$ was set to 0.2 for DNN-IVA and 1.0 for DNN-MVDR.
When training DNN-IVA, three of six channels were used and the number of iterations was 15.
DNN-MVDR was trained using all six channels.
MUSIC and Gauss-IVA also used six channels.
We used different STFT parameters for WPE and separation.
The window/shift size were set to 512/128 and 4096/1024, respectively.
The number of iterations, the delay and the tap length of WPE was set to 3, 3 and 10.
In test of IVA, we evaluated the performance using two, three and six channels with 30, 25 and 15 iterations.
When using more than two channels, two separated signals with the highest power were evaluated.
For testing, we used the averaged model parameters of the five epochs with lowest unsupervised loss on the validation set.
\\
\textbf{LibriCSS}:
We used LibriCSS~\cite{libricss} to evaluate the performance on real-recorded speeches.
Utterances taken from LibriSpeech test clean set were played back from loudspeakers and recorded by a seven-channel circular microphones with radius of 4.25~cm.
LibriCSS had 10 hours of audio recordings and they were separated into 10 sessions.
Each session contained six mini-sessions with different overlap ratios ranging from 0 to 40\%, 0S, 0L, 10, 20, 30 and 40, where 0S/0L are no overlap recordings with short/long inter-utterance silence.
Separation performance was evaluated by WERs using the ASR systems given in~\cite{libricss}, where we conducted \textit{utterance-wise} evaluation.
We used sessions 1 to 9 as training set, and session 0 as test set.

Training setup was basically the same as that on WSJ1-mix.
The difference was that BSS used only three channels to generate pseudo-targets, because using all seven channels led to over-separation, resulting in degradation of WER.
DNN-IVA also used only three channels both in training and test.
Note that MUSIC used all 7 channels to estimate DOAs, where we set the window size $L$ to 15, the shift $S$ to 1, the number of clusters $K$ to 3, the azimuth threshold $\theta_{thres}$ to 10 and the threshold to remove the outlier clusters $E_{thres}$ to $0.1\sum_{k} {E_k}$.
When training with KLD or SDR loss only, the learning rate was $10^{-5}$; otherwise $5\times 10^{-5}$.

\subsection{Results on WSJ1-mix dataset}
\label{subsec:result_wsj}
Table~\ref{table:iva_result_wsj} shows the evaluation results of IVA algorithms on the WSJ1-mix test set. 
The evaluation metrics are SDR~\cite{vincentPerformanceMeasurementBlind2006,sdrmedium}, the short-time objective intelligibility (STOI)~\cite{stoi}, the perceptual evaluation of speech quality (PESQ)~\cite{pesq} and WER.
The performance of supervised learning was also evaluated as an upper bound for unsupervised learning.
At all channel numbers, DNN-IVA trained with our proposed DOA loss outperformed conventional Gauss-IVA and DNN-IVA trained with KLD or SDR loss.
In terms of WER, our proposed method achieved comparable or even better performance than supervised learning.
We conjecture this significant WER reduction to be due to the fact that our spatial loss gives the distortionless constraint to the target source direction as in MVDR.
In addition, it is less susceptible to inter-frequency permutation problem, because the DOA $\bm{q}_{n}$ is the same regardless of  frequency.
Furthermore, the spatial loss focuses on the direct signal, which would make the model more robust against reverberation.
When we used both the spatial and the signal loss, no significant improvement was confirmed.
This implies that the spatial information is sufficient to learn the source model in IVA.

Table~\ref{table:mvdr_result_wsj} shows the evaluation results of DNN-MVDR.
Although the DOA loss worked well for IVA, it gave inferior performance to the SDR loss for MVDR.
This would be due to the fact that IVA estimates the demixing matrix so that each separated signal is independent, whereas MVDR independently estimates the demixing matrix for each source.
Thus, only the spatial information was not enough to train the network.
However, we found that combining DOA and SDR loss in learning was better than either individually.

\subsection{Results on LibriCSS dataset}
\label{subsec:result_libricss}

Table~\ref{table:result_libricss} shows the WERs of separated signals from LibriCSS dataset.
WERs per overlap ratio and average of them are listed.
Compared to the signals losses, our proposed spatial loss led to higher performance.
Both SDR and KLD loss gave only the comparable performance to pseudo-teacher, i.e., Gauss-IVA.
The poor performance on overlap-free data implies that the signals loss leads to over-separation.
Compared to DNN-IVA trained with WSJ1-mix, which contained trained ten times more data than LibriCSS, our proposed method achieved higher performance.
In addition, our proposed method also outperformed Conformer-based MVDR beamforming~\cite{libricss_conformer} with less than one twentieth the amount of data.
Furthermore, the performance was further improved by fine-tuning the supervised trained model on WSJ1-mix to LibriCSS using the proposed unsupervised learning.
Although our proposed method did not exceed state-of-the-art performance~\cite{libricss_sota}, we showed high performance with the small amount of data, which demonstrates the effectiveness of unsupervised learning with in-domain data.

\section{Conclusions}
We proposed a spatial loss function that utilizes the DOAs estimated by classic techniques.
It trains a neural separator so that it directs beam to the target source direction and null to the interference direction.
We evaluated two neural separators, DNN-IVA and DNN-MVDR, using synthetic mixtures and real-recorded LibriCSS.
In experiments with synthetic mixtures, we showed that the spatial loss worked well especially for DNN-IVA, which led to the same performance as the supervised loss.
It also performed well for DNN-MVDR when combined with the signal loss.
The spatial loss also gave high performance on LibriCSS.
It outperformed the strong baselines with the small amount of training data without any labels.

\pagebreak

\bibliographystyle{IEEEtran}

\bibliography{ver2}

\end{document}